\documentclass[10pt,a4paper]{article}
\usepackage{amsmath}
\usepackage{amssymb}
\usepackage{amsfonts}
\usepackage{amstext}
\usepackage{amsbsy}
\usepackage[mathscr]{eucal}
\usepackage{graphicx}
\usepackage{color}
\newcommand{\beq}{\begin{equation}}
\newcommand{\eeq}{\end{equation}}
\newcommand{\beqa}{\begin{eqnarray}}
\newcommand{\eeqa}{\end{eqnarray}}
\newcommand{\ket}[1]{| #1 \rangle}
\newcommand{\bra}[1]{\langle #1 |}

\makeindex
\title{\Large\textbf{Generating set for general multipartite entangled states}}

\author{\textit{ Hoshang Heydari}\\
        \small\textit{Institute of Quantum
Science, Nihon University,}\\
\small\textit{1-8 Kanda-Surugadai, Chiyoda-ku, Tokyo 101-8308, Japan
}}

\date{}
\begin{document}

\maketitle \thispagestyle{empty}

\begin{abstract}
We propose a entanglement generating set for a general multipartite
state based on  the of concurrence. In particular, we show that
concurrence for general multipartite states can be constructed by
different classes of local operators which are defined by complement
of positive operator valued measure on quantum phases. The
entanglement generating set consists of different classes of
entanglement that are detected by these classes of operators and
contributes to the degree of entanglement for a general multipartite
state.
\end{abstract}

\section{Introduction}
Quantification and classification of multipartite quantum entangled
states is an ongoing research activity in the emerging field of
quantum information and quantum computation
\cite{Lewen00,Hor00,Bennett96a,Dur00,Eisert01,Pan,Verst,Miyake}.
There are several well-known entanglement measures for bipartite
states and among these entanglement measures the concurrence
\cite{Wootters98} is widely known. Recently, there have been some
proposals to generalize this measure to general bipartite and
multipartite states \cite{Uhlmann00,Gerjuoy,Albeverio,Mintert}. We
have also defined concurrence classes for multi-qubit mixed states
and for general pure multipartite states  based on an orthogonal
complement of a positive operator valued measure ( POVM) on quantum
phase \cite{Hosh5, Hosh6}.
 In this paper, we propose a minimal
entanglement generating set (MEGS) for a general multipartite state
based on  our concurrence and it's generating operators
\cite{Hosh7}. In particular, in section \ref{povm} we give short
introduction our POVM and the construction of the operators that
generate the concurrence for general multipartite states. In section
\ref{megs} we will define a minimal entanglement generating set for
multipartite states and we will also give some example our
construction for bi-, three-, and four-partite states.

We will denote a general, composite quantum system with $m$
subsystems as $\mathcal{Q}=
\mathcal{Q}_{1}\mathcal{Q}_{2}\cdots\mathcal{Q}_{m}$ with th pure
state
\begin{equation}
\ket{\Psi}=\sum^{N_{1}}_{i_{1}=1}\cdots\sum^{N_{m}}_{i_{m}=1}
\alpha_{i_{1},i_{2},\ldots,i_{m}} \ket{i_{1},i_{2},\ldots,i_{m}}
\end{equation}
and corresponding  Hilbert space $
\mathcal{H}_{\mathcal{Q}}=\mathcal{H}_{\mathcal{Q}_{1}}\otimes
\mathcal{H}_{\mathcal{Q}_{2}}\otimes\cdots\otimes\mathcal{H}_{\mathcal{Q}_{m}},
$ where the dimension of the $j$th Hilbert space is given  by
$N_{j}=\dim(\mathcal{H}_{\mathcal{Q}_{j}})$. In order to simplify
our presentation, we will use $\Lambda_{m}=k_{1},l_{1};$
$\ldots;k_{m},l_{m}$ as an abstract
multi-index notation.   
\section{Construction of concurrence for general multipartite states}\label{povm}
In this section we give a short review  of our POVM and the
construction of concurrence for general multipartite states based on
orthogonal complement of these operators. A general and symmetric
POVM in a single $N_{j}$-dimensional Hilbert space
$\mathcal{H}_{\mathcal{Q}_{j}}$ is given by
\begin{eqnarray}
&&\Delta(\varphi_{1_j,2_j},\ldots,\varphi_{1_j,N_j},
\varphi_{2_j,3_j},\ldots,\varphi_{N_{j}-1,N_{j}})=
\sum^{N_{j}}_{l_{j},k_{j}=1}
e^{i\varphi_{k_{j},l_{j}}}\ket{k_{j}}\bra{l_{j}}\\\nonumber&&=
   \left(%
\begin{array}{ccccc}
  1 &e^{i\varphi_{1,2}}  & \cdots
  &  e^{i\varphi_{1,N_{j}-1}} &e^{i\varphi_{1,N_{j}}}\\
 e^{-i\varphi_{1,2}} &  1 & \cdots
 &  e^{i\varphi_{2,N_{j}-1}} &e^{i\varphi_{2,N_{j}}}\\
  \vdots&  \vdots&\ddots &\vdots& \vdots\\
    e^{-i\varphi_{1,N_{j}-1}} & e^{-i\varphi_{2,N_{j}-1}} &\cdots&1&e^{i\varphi_{N_{j}-1,N_{j}}}\\
  e^{-i\varphi_{1,N_{j}}} & e^{-i\varphi_{2,N_{j}}} &\cdots&e^{-i\varphi_{N_{j}-1,N_{j}}}&1\\
\end{array}%
\right),
\end{eqnarray}
where $\ket{k_{j}}$ and $\ket{l_{j}}$ are the basis vectors in
$\mathcal{H}_{\mathcal{Q}_j}$ and the quantum phases satisfy the
following relation $ \varphi_{k_{j},l_{j}}=
-\varphi_{l_{j},k_{j}}(1-\delta_{k_{j} l_{j}})$. The POVM is a
function of the $N_{j}(N_{j}-1)/2$ phases
$(\varphi_{1_j,2_j},\ldots,\varphi_{1_j,N_j},\varphi_{2_j,3_j},\ldots,\varphi_{N_{j}-1,N_{j}})$.
Moreover, our POVM is self-adjoint, positive, and  normalized. In
following text we will use the short notation
$\widetilde{\Delta}(\varphi_{k_{j},l_{j}})$  for our POVM. It is now
possible to form a POVM of a multipartite system by simply forming
the tensor product
\begin{eqnarray}\label{POVM}
\Delta_\mathcal{Q}(\varphi_{k_{1},l_{1}},\ldots,
\varphi_{k_{m},l_{m}})&=&
\Delta_{\mathcal{Q}_{1}}(\varphi_{k_{1},l_{1}}) \otimes\cdots
\otimes\Delta_{\mathcal{Q}_{m}}(\varphi_{k_{m},l_{m}}),
\end{eqnarray}
where, e.g., $\varphi_{k_{1},l_{1}}$ is the set of POVMs phase
associated with subsystems $\mathcal{Q}_{1}$, for all
$k_{1},l_{1}=1,2,\ldots,N_{1}$.
In the $m$-partite case, the off-diagonal elements of the matrix
corresponding to
\begin{equation}\widetilde{\Delta}_\mathcal{Q}(\varphi_{k_{1},l_{1}},\ldots,
\varphi_{k_{m},l_{m}})=
\widetilde{\Delta}_{\mathcal{Q}_{1}}(\varphi_{k_{1},l_{1}})
\otimes\cdots
\otimes\widetilde{\Delta}_{\mathcal{Q}_{m}}(\varphi_{k_{m},l_{m}}),
\end{equation}
 where the orthogonal complement of our POVM  is
defined by
$\widetilde{\Delta}(\varphi_{k_{j},l_{j}})=\mathcal{I}_{N_{j}}-
\Delta_{\mathcal{Q}_{j}}(\varphi_{k_{j},l_{j}})$.
$\mathcal{I}_{N_{j}}$ is the $N_{j}$-by-$N_{j}$ identity matrix for
subsystem $j$.

 $\widetilde{\Delta}_\mathcal{Q}(\varphi_{k_{1},l_{1}},\ldots,
\varphi_{k_{m},l_{m}})$ has phases that are sums or differences of
phases originating from  $2,3,\ldots,m$ subsystems. That is, in the
latter case the phases of
$\widetilde{\Delta}_\mathcal{Q}(\varphi_{k_{1},l_{1}},\ldots,
\varphi_{\mathcal{Q}_{m};k_{m},l_{m}})$ take the form
$(\varphi_{k_{1},l_{1}}\pm\varphi_{k_{2},l_{2}}
\pm\ldots\pm\varphi_{k_{m},l_{m}})$ and identification of these
joint phases makes our distinguishing possible. Thus, we can define
linear operators for the $\mathrm{EPR}^{m}$ class which are sums and
differences of phases of two subsystems, i.e.,
$(\varphi_{k_{r_{1}},l_{r_{1}}} \pm\varphi_{k_{r_{2}},l_{r_{2}}})$.
That is, for the $\mathrm{EPR}^{m}$ class we have
\begin{equation}
 \widetilde{\Delta}^{
\mathrm{EPR}^{m}_{\Lambda_{m}}}_{\mathcal{Q}_{r_{1},r_{2}}(N_{r_{1}},N_{r_{2}})}
=\mathcal{I}_{N_{1}} \otimes\cdots
\otimes\widetilde{\Delta}_{\mathcal{Q}_{r_{1}}}
(\varphi^{\frac{\pi}{2}}_{k_{r_{1}},l_{r_{1}}}) \otimes\cdots\otimes
\widetilde{\Delta}_{\mathcal{Q}_{r_{2}}}
(\varphi^{\frac{\pi}{2}}_{k_{r_{2}},l_{r_{2}}})\otimes\cdots\otimes\mathcal{I}_{N_{m}},
\end{equation}
where $\varphi^{\frac{\pi}{2}}_{k_{j},l_{j}}=\frac{\pi}{2}$ for all
$k_{j}<l_{j}, ~j=1,2,\ldots,m$.
Next, we rewrite the linear operator
$\widetilde{\Delta}^{\mathrm{EPR}^{m}_{\Lambda_{m}}}_{\mathcal{Q}_{r_{1},r_{2}}(N_{r_{1}},N_{r_{2}})}$
as a direct sum of the upper and lower anti-diagonal
\begin{equation}
 \widetilde{\Delta}^{
\mathrm{EPR}^{m}_{\Lambda_{m}}}_{\mathcal{Q}_{r_{1},r_{2}}(N_{r_{1}},N_{r_{2}})}
=U\widetilde{\Delta}^{
\mathrm{EPR}^{m}_{\Lambda_{m}}}_{\mathcal{Q}_{r_{1},r_{2}}(N_{r_{1}},N_{r_{2}})}+L\widetilde{\Delta}^{
\mathrm{EPR}^{m}_{\Lambda_{m}}}_{\mathcal{Q}_{r_{1},r_{2}}(N_{r_{1}},N_{r_{2}})}.
\end{equation}
 The  set of linear operators for the $\mathrm{EPR}^{m}$ classes
gives
the $\mathrm{W}^{m}$ class concurrence. 
The next class we will consider is what we call the
$\mathrm{GHZ}^{m}$ class which given by
\begin{eqnarray}\nonumber
 \widetilde{\Delta}^{
\mathrm{GHZ}^{m}_{\Lambda_{m}}}_{\mathcal{Q}_{r_{1},r_{2}}(N_{r_{1}},N_{r_{2}})}
&=&\widetilde{\Delta}_{\mathcal{Q}_{1}}
(\varphi^{\pi}_{k_{1},l_{1}})\otimes\cdots
\otimes\widetilde{\Delta}_{\mathcal{Q}_{r_{1}}}
(\varphi^{\frac{\pi}{2}}_{k_{r_{1}},l_{r_{1}}}) \otimes\cdots\otimes
\widetilde{\Delta}_{\mathcal{Q}_{r_{2}}}
(\varphi^{\frac{\pi}{2}}_{k_{r_{2}},l_{r_{2}}})\\&&\otimes\cdots\otimes
\widetilde{\Delta}_{\mathcal{Q}_{m}} (\varphi^{\pi}_{k_{m},l_{m}}),
\end{eqnarray}
 where by choosing $\varphi^{\pi}_{k_{j},l_{j}}=\pi$ for all
$k_{j}<l_{j}, ~j=1,2,\ldots,m$, we get an operator which has the
structure of the Pauli operator $\sigma_{x}$ embedded in a
higher-dimensional Hilbert space and coincides with $\sigma_{x}$ for
a single-qubit. 
Next, we write the linear operators for the $\mathrm{GHZ}^{m}$ class
as
\begin{equation}
 \widetilde{\Delta}^{
\mathrm{GHZ}^{m}_{\Lambda_{m}}}_{\mathcal{Q}_{r_{1},r_{2}}(N_{r_{1}},N_{r_{2}})}
=P_{1}\widetilde{\Delta}^{
\mathrm{GHZ}^{m}_{\Lambda_{m}}}_{\mathcal{Q}_{r_{1},r_{2}}(N_{r_{1}},N_{r_{2}})}+P_{2}\widetilde{\Delta}^{
\mathrm{GHZ}^{m}_{\Lambda_{m}}}_{\mathcal{Q}_{r_{1},r_{2}}(N_{r_{1}},N_{r_{2}})}+\ldots,
\end{equation}

 where the
operators $P_{i}\widetilde{\Delta}^{
\mathrm{GHZ}^{m}_{\Lambda_{m}}}_{\mathcal{Q}_{r_{1},r_{2}}(N_{r_{1}},N_{r_{2}})}$
are constructed by pairing of elements of the POVM with sums and
differences of quantum phases. 
We can also define the $\mathrm{GHZ}^{m-1}$ class operator
\begin{eqnarray} \widetilde{\Delta}^{
\mathrm{GHZ}^{m-1}_{\Lambda_{m}}}_{\mathcal{Q}_{r_{1}r_{2},r_{3}}(N_{r_{1}},N_{r_{2}})}
&=&\widetilde{\Delta}_{\mathcal{Q}_{r_{1}}}
(\varphi^{\frac{\pi}{2}}_{k_{r_{1}},l_{r_{1}}})
\otimes\widetilde{\Delta}_{\mathcal{Q}_{r_{2}}}
(\varphi^{\frac{\pi}{2}}_{k_{r_{2}},l_{r_{2}}})
\otimes\widetilde{\Delta}_{\mathcal{Q}_{r_{3}}}
(\varphi^{\pi}_{k_{r_{3}},l_{r_{3}}}) \\\nonumber&&\otimes\cdots
\otimes\widetilde{\Delta}_{\mathcal{Q}_{m-1}}
(\varphi^{\pi}_{k_{r_{m-1}},l_{r_{m-1}}})\otimes\mathcal{I}_{N_{m}}
,
\end{eqnarray}
 where $1\leq r_{1}<r_{2}<\cdots<r_{m-1}<m$. In the same way we can
construct all other operators in these classes. Now, by taking the
expectation value of each of these classes of operators, we are able
to construct the concurrence for general multipartite states
\cite{Hosh7}. In the next section we will propose a minimal
entanglement generating set for general multipartite states  based
on our construction of the concurrence.
\section{Minimal entanglement generating set for multipartite
states}\label{megs} As we have shown there are different classes of
operators that generates concurrence for general multipartite
states. So, the construction of a measures of entanglement for
general multipartite states suggests that there exists different
classes of entanglement that contributes to the degree entanglement.
These different classes of entanglement are detected by
$\mathrm{EPR}^{m}$ and $\mathrm{GHZ}^{m}$ classes of operators.
Thus, we can propose a $\mathrm{MEGS}$ for general multipartite
states based on this construction of the concurrence as follows
\begin{eqnarray}\label{MEGS}\nonumber
\mathcal{E}^{m}_{MEGS}&=&\{\mathrm{EPR}_{\mathcal{Q}_{1}\mathcal{Q}_{2}}
,\ldots,\mathrm{EPR}_{\mathcal{Q}_{1}\mathcal{Q}_{m}},\ldots,
\mathrm{EPR}_{\mathcal{Q}_{m-2}\mathcal{Q}_{m-1}},
\ldots,\mathrm{EPR}_{\mathcal{Q}_{m-1}\mathcal{Q}_{m}},\\\nonumber&&
\mathrm{GHZ}^{3}_{\mathcal{Q}_{1}\mathcal{Q}_{2}\mathcal{Q}_{3}},\ldots
\mathrm{GHZ}^{3}_{\mathcal{Q}_{m-2}\mathcal{Q}_{m-1}\mathcal{Q}_{m}},\ldots,
,
\mathrm{GHZ}^{m-1}_{\mathcal{Q}_{1}\mathcal{Q}_{2}\cdots\mathcal{Q}_{m-1}},\ldots,\\&&
\mathrm{GHZ}^{m-1}_{\mathcal{Q}_{2}\mathcal{Q}_{3}\cdots\mathcal{Q}_{m}},
\mathrm{GHZ}^{m}_{\mathcal{Q}_{1}\mathcal{Q}_{2}\cdots\mathcal{Q}_{m}}\},
\end{eqnarray}
where for $m$-partite states we have $C(m,2)=\frac{m(m-1)}{2}$
$\mathrm{EPR}_{\mathcal{Q}_{r_{1}}\mathcal{Q}_{r_{2}}}$ classes and
$C(m,k)$
~$\mathrm{GHZ}^{k}_{\mathcal{Q}_{r_{1}}\mathcal{Q}_{r_{2}}\cdots\mathcal{Q}_{r_{k}}}$
classes, for $2<k\leq m$.
For example the minimal entanglement generating set for bipartite
states has only one element, that is
$\mathcal{E}^{2}_{MEGS}=\{\mathrm{EPR}_{\mathcal{Q}_{1}\mathcal{Q}_{2}}
 \} $. This is exactly what we have expect to see. For three-partite states
 the MEGS has $C(3,2)=3$ EPR elements  and $C(3,3)=1$ $\mathrm{GHZ}^{3}$ element
\begin{equation}
\mathcal{E}^{3}_{MEGS}=\{\mathrm{EPR}_{\mathcal{Q}_{1}\mathcal{Q}_{2}}
,\mathrm{EPR}_{\mathcal{Q}_{1}\mathcal{Q}_{3}},\mathrm{EPR}_{\mathcal{Q}_{2}\mathcal{Q}_{3}},
\mathrm{EPR}_{\mathcal{Q}_{1}\mathcal{Q}_{2}\mathcal{Q}_{3}} \}.
\end{equation}
Note also that the combinations of EPR elements gives the W class
entanglement. Our last example is the MEGS for four-partite states
which is given by
\begin{eqnarray}
\mathcal{E}^{4}_{MEGS}&=&\{\mathrm{EPR}_{\mathcal{Q}_{1}\mathcal{Q}_{2}}
,\mathrm{EPR}_{\mathcal{Q}_{1}\mathcal{Q}_{3}},\mathrm{EPR}_{\mathcal{Q}_{1}\mathcal{Q}_{4}},
\mathrm{EPR}_{\mathcal{Q}_{2}\mathcal{Q}_{3}},
\mathrm{EPR}_{\mathcal{Q}_{2}\mathcal{Q}_{4}}
,\\\nonumber&&\mathrm{EPR}_{\mathcal{Q}_{3}\mathcal{Q}_{4}},
\mathrm{GHZ}^{3}_{\mathcal{Q}_{1}\mathcal{Q}_{2}\mathcal{Q}_{3}},
\mathrm{GHZ}^{3}_{\mathcal{Q}_{1}\mathcal{Q}_{2}\mathcal{Q}_{4}},
\mathrm{GHZ}^{3}_{\mathcal{Q}_{1}\mathcal{Q}_{3}\mathcal{Q}_{4}},
\mathrm{GHZ}^{3}_{\mathcal{Q}_{2}\mathcal{Q}_{3}\mathcal{Q}_{4}},\\\nonumber&&
\mathrm{GHZ}^{4}_{\mathcal{Q}_{1}\mathcal{Q}_{2}\mathcal{Q}_{3}\mathcal{Q}_{4}}\}.
\end{eqnarray}
This set has $C(4,2)=6$ EPR elements, $C(4,3)=4$ $\mathrm{GHZ}^{3}$
elements, and $C(4,4)=1$ $\mathrm{GHZ}^{4}$ element. Note that, for
each element of EPR class there is only one operator that correspond
to a given element of the MEGS. But for each element of GHZ classes
there are a set of operators that correspond to a given element of
the MEGS. The elements of MEGS are inequivalent under local quantum
operations and classical communication by construction an in the
case of three-partite states and in general between
$\mathrm{EPR}^{m}$ and $\mathrm{GHZ}^{m}$ classes.
 The MEGS set is not equivalent to minimal reversible
entanglement generating set, but there is some similarity between
these two sets.
 Thus the MEGS set
 gives
 information
about the nature of multipartite entangled states in a time where
there is no well known and accepted classification of multipartite
states available.
\begin{flushleft}
\textbf{Acknowledgments:} The  author  acknowledges the financial
support of the Japan Society for the Promotion of Science (JSPS).
\end{flushleft}


\end{document}